\newif\if@defeqnsw \@defeqnswtrue

\def\eqnarray{\stepcounter{equation}\let\@currentlabel=\theequation
\if@defeqnsw\global\@eqnswtrue\else\global\@eqnswfalse\fi
\global\@eqnswtrue
\tabskip\@centering\let\\=\@eqncr
$$\halign to \displaywidth\bgroup\hfil\global\@eqcnt\z@
  $\displaystyle\tabskip\z@{##}$&\global\@eqcnt\@ne
  \hfil$\displaystyle{{}##{}}$\hfil
  &\global\@eqcnt\tw@ $\displaystyle{##}$\hfil
  \tabskip\@centering&\llap{##}\tabskip\z@\cr}

\def\yesnumber{\global\@eqnswtrue}

\def\@@eqncr{\let\@tempa\relax\global\advance\@eqcnt by \@ne
    \ifcase\@eqcnt \def\@tempa{& & & &}\or \def\@tempa{& & &}\or
     \def\@tempa{& &}\or \def\@tempa{&}\else\fi
     \@tempa \if@eqnsw\@eqnnum\stepcounter{equation}\fi
     \if@defeqnsw\global\@eqnswtrue\else\global\@eqnswfalse\fi
     \global\@eqcnt\z@\cr}

\def\@eqnacr{{\ifnum0=`}\fi\@ifstar{\@yeqnacr}{\@yeqnacr}}

\def\@yeqnacr{\@ifnextchar [{\@xeqnacr}{\@xeqnacr[\z@]}}

\def\@xeqnacr[#1]{\ifnum0=`{\fi}\cr \noalign{\vskip\jot\vskip #1\relax}}

\def\eqalign{\null\,\vcenter\bgroup\openup1\jot \m@th \let\\=\@eqnacr
\ialign\bgroup\strut
\hfil$\displaystyle{##}$&$\displaystyle{{}##}$\hfil\crcr}
\def\endeqalign{\crcr\egroup\egroup\,}

\def\cases{\left\{\,\vcenter\bgroup\normalbaselines\m@th \let\\=\@eqnacr
    \ialign\bgroup$##\hfil$&\quad##\hfil\crcr}
\def\endcases{\crcr\egroup\egroup\right.}

\def\eqalignno{\stepcounter{equation}\let\@currentlabel=\theequation
\if@defeqnsw\global\@eqnswtrue\else\global\@eqnswfalse\fi
\let\\=\@eqncr
$$\displ@y \tabskip\@centering \halign to \displaywidth\bgroup
  \global\@eqcnt\@ne\hfil
  $\@lign\displaystyle{##}$\tabskip\z@skip&\global\@eqcnt\tw@
  $\@lign\displaystyle{{}##}$\hfil\tabskip\@centering&
  \llap{\@lign##}\tabskip\z@skip\crcr}

\def\endeqalignno{\@@eqncr\egroup
      \global\advance\c@equation\m@ne$$\global\@ignoretrue}

\@namedef{eqalignno*}{\@defeqnswfalse\eqalignno}
\@namedef{endeqalignno*}{\endeqalignno}

\def\eqaligntwo{\stepcounter{equation}\let\@currentlabel=\theequation
\if@defeqnsw\global\@eqnswtrue\else\global\@eqnswfalse\fi
\let\\=\@eqncr
$$\displ@y \tabskip\@centering \halign to \displaywidth\bgroup
  \global\@eqcnt\m@ne\hfil
  $\@lign\displaystyle{##}$\tabskip\z@skip&\global\@eqcnt\z@
  $\@lign\displaystyle{{}##}$\hfil\qquad&\global\@eqcnt\@ne
  \hfil$\@lign\displaystyle{##}$&\global\@eqcnt\tw@
  $\@lign\displaystyle{{}##}$\hfil\tabskip\@centering&
  \llap{\@lign##}\tabskip\z@skip\crcr}

\def\endeqaligntwo{\@@eqncr\egroup
      \global\advance\c@equation\m@ne$$\global\@ignoretrue}

\@namedef{eqaligntwo*}{\@defeqnswfalse\eqaligntwo}
\@namedef{endeqaligntwo*}{\endeqaligntwo}

\newtoks\@stequation

\def\subequations{\refstepcounter{equation}%
  \edef\@savedequation{\the\c@equation}%
  \@stequation=\expandafter{\theequation}%   %only want \theequation
  \edef\@savedtheequation{\the\@stequation}% %expanded once
  \edef\oldtheequation{\theequation}%
  \setcounter{equation}{0}%
  \def\theequation{\oldtheequation\alph{equation}}}

\def\endsubequations{%
  \setcounter{equation}{\@savedequation}%
  \@stequation=\expandafter{\@savedtheequation}%
  \edef\theequation{\the\@stequation}%
  \global\@ignoretrue}

\def\big#1{{\hbox{$\left#1\vcenter to1.428\ht\strutbox{}\right.\n@space$}}}
\def\Big#1{{\hbox{$\left#1\vcenter to2.142\ht\strutbox{}\right.\n@space$}}}
\def\bigg#1{{\hbox{$\left#1\vcenter to2.857\ht\strutbox{}\right.\n@space$}}}
\def\Bigg#1{{\hbox{$\left#1\vcenter to3.571\ht\strutbox{}\right.\n@space$}}}

\endinput

\documentstyle[12pt,equation]{article}
\begin{document}
\newcommand{\aq}{\mbox{$a_{\tilde{Q}}$}}
\newcommand{\bq}{\mbox{$b_{\tilde{Q}}$}}
\newcommand{\asq}{\mbox{$a^2_{\tilde{Q}}$}}
\newcommand{\bsq}{\mbox{$b^2_{\tilde{Q}}$}}
\newcommand{\tanb}{\mbox{$\tan \! \beta$}}
\newcommand{\mlsq}{\mbox{$m^2_{H_2}$}}
\newcommand{\mhsq}{\mbox{$m^2_{H_1}$}}
\newcommand{\mx}{\mbox{$M_X$}}
\newcommand{\rt}{\mbox{$\sqrt{|\Delta|}$}}
\newcommand{\mc}{\mbox{$m_{\chi}$}}
\newcommand{\mcsq}{\mbox{$m^2_{\chi}$}}
\newcommand{\msq}{\mbox{$m_{\tilde{q}}$}}
\newcommand{\msqsq}{\mbox{$m^2_{\tilde{q}}$}}
\newcommand{\msQ}{\mbox{$m_{\tilde{Q}}$}}
\newcommand{\mQsq}{\mbox{$m^2_Q$}}
\newcommand{\msQsq}{\mbox{$m^2_{\tilde{Q}}$}}
\newcommand{\msbsq}{\mbox{$m^2_{\tilde{b}_1}$}}
\newcommand{\msb}{\mbox{$m_{\tilde{b}_1}$}}
\newcommand{\mstau}{\mbox{$m_{\tilde{\tau}_1}$}}
\newcommand{\mst}{\mbox{$m_{\tilde{t}_1}$}}
\newcommand{\mstsq}{\mbox{$m^2_{\tilde{t}_1}$}}
\newcommand{\gsq}{\frac {g_s^2}{16 \pi^2}}
\newcommand{\be}{\begin{equation}}
\newcommand{\ee}{\end{equation}}
\newcommand{\een}{\end{subequations}}
\newcommand{\ben}{\begin{subequations}}
\newcommand{\beq}{\begin{eqalignno}}
\newcommand{\eeq}{\end{eqalignno}}
\renewcommand{\thefootnote}{\fnsymbol{footnote} }
\noindent
\begin{flushright}
MAD/PH/723\\
October 1992
\end{flushright}
\vspace{1.5cm}
\begin{center}
{\Large \bf New Contributions to Coherent Neutralino--Nucleus Scattering}\\
\vspace{5mm}
Manuel Drees\\
{\em Theorie-Gruppe, DESY, Notkestr. 85, D2000 Hamburg 52, Germany} \\
\vspace{5mm}
Mihoko M. Nojiri\footnote{e-mail address; PHENOC::NOJIRI, NOJIRI@WISCPHEN}\\
{\em Physics Department, University of Wisconsin, Madison, WI 53706, USA}
\end{center}

\begin{abstract}

{\small We discuss coherent scattering between supersymmetric neutralinos and
nuclei via an effective neutralino--gluon interaction. We identify two new
classes of diagrams which are not treated in the existing literature. These
occur at the same order of perturbation theory as the previously considered
diagrams. The new contributions can be numerically important, and can even
dominate the total amplitude. This affects the calculation of event rates in
experiments searching for direct or indirect evidence of cosmic relic
neutralinos.}

\end{abstract}

\clearpage
\noindent

\pagestyle{plain}
\setcounter{page}{1}
One of the attractive features of many supersymmetric (SUSY) models \cite{1} is
that they offer an explanation for the observed \cite{2} Dark Matter (DM) in
the Universe: In models with ``R-parity'' conservation, the lightest
supersymmetric particle (LSP) is  stable; moreover, it has been shown in
detailed model calculations \cite{3} that in wide regions of parameter space
the relic density of LSPs left over from the Big Bang naturally has a
cosmologically interesting value. It should be emphasized here that the main
motivation for the introduction of supersymmetry is \cite{4} the stabilization
of the huge hierarchy between the weak scale on the one hand and the Grand
Unified (GUT) or Planck scale on the other. The fact that SUSY can {\em in
addition} solve a long--standing cosmological problem makes the LSP an
especially interesting DM candidate. It is also worth mentioning that precision
LEP data favour \cite{GUT} a supersymmetric Grand Unification over the
conventional one.

A stable LSP has to be electrically and color neutral, since otherwise it
would have been detected in searches \cite{5} for exotic isotopes. Within the
supersymmetric version of the Standard Model, the MSSM, this leaves us with
two candidates: The sneutrino $\tilde{\nu}$, and the lightest neutralino
$\chi$. The former interacts with ordinary matter with full weak interaction
strength (via $Z$ exchange); the first round of direct DM detection
experiments \cite{6} already reached sufficient sensitivity to exclude the
sneutrino as a sizeable component of the DM halo of our galaxy. This
negative result is actually not surprising, since in most models \cite{1} one
ha
s
$m_{\tilde{\nu}} > \mc$, i.e. the sneutrino is {\em not} the LSP.

Unfortunately, relic neutralinos are much more difficult to detect than
relic sneutrinos (or heavy relic Dirac neutrinos). In the limit where $\chi$
is a pure higgsino or gaugino, its interactions with matter are suppressed
by a factor $(M_Z/\msq)^4 \leq 0.1$, compared to the case of sneutrinos.
Moreover, in lowest order it interacts with quarks only via an axial--vector
current \cite{7}; the corresponding hadronic matrix element involves the
spin of the nucleon.\footnote{Unless there is sizable mixing between the
superpartners of left-- and right--handed quarks \cite{7,8}; however, this
mixing is expected \cite{1} to be of the order $m_q/\msq$, and thus
negligible for the superpartners of quarks light enough to be found in the
nucleon with non--negligible probability.} This means that the scattering
matrix element off a nucleus will {\em not} be larger than that off a
nucleon, and might even be smaller \cite{9}. In contrast,
spin--independent interactions lead to a coherent LSP--nucleus coupling,
so that the matrix element is proportional to
the number of nucleons inside that nucleus. Therefore a small spin--independent
contribution to the LSP--nucleon scattering matrix element can lead to a
large enhancement of the LSP--nucleus cross section, and thus of the
counting rate in a given DM detector.

In existing calculations \cite{9,10} of LSP detection rates it has been assumed
that the spin--independent contribution to the LSP--nucleon cross section is
dominated by diagrams with scalar Higgs boson exchange. The Higgs bosons couple
both to light ($u,d,s$) quarks, whose (current) masses and nucleonic matrix
elements $\langle N | q \bar{q} | N \rangle$ have to be taken from
non--perturbative model estimates \cite{11}, and to heavy quarks, whose
abundance ``in'' the nucleon can be calculated perturbatively \cite{12}. This
latter contribution can also be understood as being due to an effective
LSP -- gluon interaction produced by the diagrams of fig.\,1a, multiplied with
the matrix element $\langle N | F_{\mu\nu} F^{\mu\nu} | N \rangle$, which
is related \cite{12} to the nucleon mass; here $F_{\mu\nu}$ is the gluon
field strength tensor. In this Letter we point out that at the same order
of perturbation theory, there exist additional contributions to the
effective LSP--gluon interaction, which in certain cases can overwhelm the
contribution from fig.\,1a.

The starting point of our calculation is the well--known observation \cite{13}
that the superpartners of heavy left-- and right--handed quarks, in
particular the top quark, mix strongly. As a result, one of the eigenstates
can be substantially lighter than the superpartners of light quarks; moreover,
the coupling of the Higgs boson to superpartners of heavy quarks can be of
the order of the squark mass. In this case, the squark triangle and bubble
diagrams of fig.\,1b can be as important as the quark triangle diagrams of
fig.\,1a. Furthermore, there are also diagrams which have nothing to do with
Higgs exchange, see fig.\,1c; in these box and triangle diagrams the LSP
splits directly into a quark and a squark, and the 2 gluons are attached in
all possible combinations to the squark and/or quark line. Notice that these
latter diagrams give a nonzero contribution even if the LSP is a pure (unmixed)
state; in contrast, the contributions from the Higgs exchange diagrams of
figs.\,1a,b will vanish unless the LSP has both gaugino and higgsino
components. Since in many realistic cases the LSP is an almost pure
state, it is conceivable that in many cases the diagrams of fig.\,1c are
{\em more} important than those of fig.\,1a. We now proceed to show that this
can indeed be the case.

The diagrams of fig.\,1a give rise to an effective Lagrangian of the form
\be \label{e1}
{\cal L}_{\rm{eff}}(1a) = T_q \bar{\chi} \chi F_{\mu\nu}^a F^{\mu\nu}_a,
\ee
where $a$ is a color index. The MSSM contains two scalar Higgs bosons;
following
the conventions of ref.\cite{14} we denote them by $H_1^0$ and $H_2^0$,
$H_1^0$ being the heavier one. Their couplings to quarks can be written as
\cite{14} $c_Q^{(i)} m_Q, \ i=1,2$; here $c_Q^{(i)}$ does not depend on the
quark mass $m_Q$, which implies that the contribution of a heavy quark to
the coefficient $T_q$ is also independent of $m_Q$. Altogether one has:
\be \label{e2}
T_q = \gsq \cdot \frac{1}{3} \cdot \sum_{i=1}^2 \frac {c^{(i)}_{\chi}}
{m^2_{H_0^i}} \sum_{Q=c,b,t} c^{(i)}_Q, \ee
where $c^{(i)}_{\chi}$ is the $H_0^i \bar{\chi} \chi$ coupling \cite{14}; as
already mentioned, $c^{(i)}_{\chi} \ne 0$ only if $\chi$ has both gaugino
and higgsino components. Notice that the $c^{(i)}_{\chi}$ are dimensionless,
while the $c^{(i)}_Q$ are $\propto 1/M_Z$.\footnote{Since the LSP is a
Majorana particle, we also have to include diagrams with crossed LSP
lines. For the diagrams of figs.\,1a,b, this simply amounts to multiplying the
result with a factor of 2. However, the same factor of 2 appears when one
writes
down the matrix element directly from the effective Lagrangian \ref{e1}; it is
therefore {\em not} included in $T_q$, eq.(\ref{e2}).}

Eq.(\ref{e2}) and our subsequent results are valid only in the limit of
small momentum transfer. Since the velocity of relic neutralinos in the
vicinity of the Earth is expected to be about 300 km/sec, the maximal possible
squared 4--momentum transmitted to the gluons is well below 1 ${\rm GeV}^2$,
and thus much smaller than all masses in the loop.

The diagrams of fig.\,1b give rise to the same kind of effective Lagrangian
as in eq.(\ref{e1}), but with coefficient \be \label{e3}
T_{\tilde{q}} = \gsq \cdot \frac{1}{24} \sum_{i=1}^2 \frac {c^{(i)}_{\chi}}
{m^2_{H_0^i}} \sum_{\tilde{Q}} \frac {c^{(i)}_{\tilde{Q}}} {m^2_{\tilde{Q}}},
\ee
where the sum over $\tilde{Q}$ runs over both eigenstates of the superpartners
of heavy quarks. The couplings $c^{(i)}_{\tilde{Q}}$ have dimension of
mass. They get \cite{14} supersymmetric ``$D$--term'' contributions $\propto
M_Z$ and ``$F$--term'' contributions $\propto \mQsq/M_Z$, as well as
contributions proportional to the Higgs(ino) mass parameter $\mu$ and the soft
breaking parameters $A_Q$. The interactions $\propto \mu$ and $A_Q$ also give
rise to mixing between left-- and right--handed squarks when the Higgs boson
is replaced by its vev:
\ben \label{em1} \beq
{\cal M}^2_{\tilde t} &= \mbox{$ \left( \begin{array}{cc}
m^2_{\tilde{t}_L} + m_t^2 + 0.35 D_Z & - m_t (A_t + \mu \cot \! \beta) \\
- m_t (A_t + \mu \cot \! \beta ) & m^2_{\tilde{t}_R} + m_t^2 + 0.16 D_Z
\end{array} \right) $}; \label{em1a} \\
{\cal M}^2_{\tilde b} &= \mbox{$ \left( \begin{array}{cc}
m^2_{\tilde{t}_L} + m_b^2 - 0.42 D_Z & - m_b (A_b + \mu \tanb) \\
- m_b (A_b + \mu \tanb) & m^2_{\tilde{b}_R} + m_b^2 - 0.08 D_Z
\end{array} \right) $}, \label{em1b} \eeq \een
where $D_Z = M_Z^2 \cos \! 2 \beta$, and the
$m^2_{\tilde{t}_L,\tilde{t}_R,\tilde{b}_R}$ are soft breaking masses which
do not enter the Higgs--squark--squark couplings. The $\tilde c$ mass matrix
is identical to the one for $\tilde t$, with $t \rightarrow c$.

The expressions for the $c^{(i)}_{\tilde{Q}}$ including squark mixing are too
lengthy to be reproduced here, but they can be obtained quite easily from
results of ref.\cite{14}.\footnote{Notice that as in ref.\cite{15}, but unlike
ref.\cite{14}, we have defined the off--diagonal elements of the squark mass
matrices to have negative sign; the same sign also appears in the
$c^{(i)}_{\tilde{Q}}$.}  We include stop and sbottom eigenstates in the loop
since the lighter sbottom eigenstate can have \cite{15} a reduced mass if the
ratio \tanb\ of the vevs of the two Higgs doublets is large, see
eq.(\ref{em1b}); moreover, the $D$--term contributions cancel after summation
over a degenerate full family of squarks, making it necessary to include
$\tilde{b}$ loops even for small \tanb.

\setcounter{footnote}{0}
The box and triangle diagrams of fig.\,1c give rise to a somewhat more
complicated effective Lagrangian: \beq \label{e4}
{\cal L}_{\rm{eff}}(1c) &= (B_D + B_S) \bar{\chi} \chi F_{\mu\nu}^a
F^{\mu\nu}_a + (B_{1D}+B_{1S}) \bar{\chi} \partial_{\mu} \partial_{\nu}
\chi F_a^{\mu \rho} F^{\nu a}_{\rho} \nonumber \\
&+ B_{2S} \bar{\chi} (i \partial_{\mu} \gamma_{\nu} + i \partial_{\nu}
\gamma_{\mu}) \chi F_a^{\mu \rho} F^{\nu a}_{\rho}. \eeq
The first two terms in this effective Lagrangian break chirality, while the
third does not. The coefficients of the chirality breaking terms have to
contain an explicit factor of a fermion mass, either the LSP mass \mc\ or
the quark mass $m_Q$; since the corresponding contributions have different
coupling structure, we have written them separately in eq.(\ref{e4}). Only the
first term contributes to the same hadronic matrix element that enters the
Higgs exchange contributions of figs.\,1a,b. The corresponding coefficients
$B_D$ and $B_S$ are given by: \ben \label{e5} \beq
B_D &= \gsq \cdot \frac {1}{8} \sum_{\tilde{Q}} (\asq - \bsq) m_Q
I_1(\msQ,m_Q,\mc); \label{e5a} \\
B_S &= \gsq \cdot \frac {1}{8} \mc \sum_{\tilde{Q}} (\asq + \bsq)
I_2(\msQ,m_Q,\mc). \label{e5b} \eeq \een
Here, \aq\ and \bq\ determine the LSP--quark--squark interaction:
\be \label{e6}
{\cal L}_{Q \tilde{Q} \chi} = \sum_{\tilde{Q}} \bar{Q} (\aq + \bq \gamma_5)
\chi \tilde{Q} + h.c.; \ee
recall that the sum over $\tilde{Q}$ includes both eigenstates of squarks with
a given flavor. In the notation of ref.\cite{16}, these couplings are given
by: \be \label{e7}
a_{\tilde{Q}_1} = \frac{1}{2} \left( X'_{Q0} + W'_{Q0} \right); \ \ \ \ \
b_{\tilde{Q}_1} = \frac{1}{2} \left( X'_{Q0} - W'_{Q0} \right), \ee
for the lighter eigenstate $\tilde{Q}_1$; the corresponding expressions for
the heavier eigenstate $\tilde{Q}_2$ can be obtained by the replacement
$W' \rightarrow Z', X' \rightarrow Y'$. Finally, the Feynman parameter
integrals
$I_1$ and $I_2$ are given by: \ben \label{e8} \beq
I_1(\msQ,m_Q,\mc) &= \int_0^1 dx \frac {x^2 - 2x + 2/3} {D^2} \\ \label{e8a}
&= \frac {1} {\Delta} \left[ \frac {\mQsq-\mcsq}{3 \msQsq} - \frac {2}{3}
\frac {\msQsq - \mcsq}{\mQsq} - \frac{5}{3} + \left(2 \msQsq - \frac{2}{3}
\mcsq \right) L \right]; \nonumber \\
I_2(\msQ,m_Q,\mc) &=  \int_0^1 dx \frac {x(x^2 - 2x + 2/3)} {D^2}
\nonumber \\
&= \frac {1}{2 m^4_{\chi}} \left[ \ln \frac {\msQsq} {\mQsq} - \left(
\msQsq -\mQsq - \mcsq \right) L \right] \label{e8b} \\
&+ \frac {1} {\Delta} \left\{ \left[ \frac {m_Q^4 - \mQsq \msQsq} {\mcsq}
- \frac {7}{3} \mQsq + \frac{2}{3} (\mcsq - \msQsq) \right] L \right.
\nonumber \\ & \left. \hspace*{1cm} + \frac {\mQsq -\mcsq} {3\msQsq}
+ \frac {\msQsq - \mQsq} {\mcsq} + \frac {2}{3} \right\}. \nonumber \eeq \een
In these expressions we have assumed $\mc < \msQ$, and
have introduced: \ben \label{e9} \beq
D &= x^2 \mcsq + x \left( \msQsq - \mQsq - \mcsq \right) + \mQsq;
\label{e9a} \\
\Delta &= 2 \mcsq \left (\mQsq + \msQsq \right) - m^4_{\chi} -
\left( \msQsq - \mQsq \right)^2; \label{e9b} \\
L &= \frac {2} {\sqrt{|\Delta|}} \arctan \frac {\rt} {\mQsq + \msQsq - \mcsq},
\ \ \ \ \ \ \Delta \geq 0, \nonumber \\
 &= \frac {1} {\sqrt{|\Delta|}} \ln \frac {\mQsq + \msQsq - \mcsq + \rt}
{\mQsq + \msQsq - \mcsq - \rt}, \ \ \
\Delta \leq 0
\label{e9c} \eeq \een

Notice that $I_1$ and $I_2$ remain finite as $\Delta \rightarrow 0$ or $\mc
\rightarrow 0$. $I_1$ diverges quadratically as $m_Q \rightarrow 0$, but $B_D$
remains finite in this limit due to the explicit factor of $m_Q$ in
eq.(\ref{e5a}), as well as the fact that $\asq - \bsq \propto m_Q$: The mixing
between left-- and right--handed squarks is proportional to the current quark
mass, so that for $m_Q = 0$ one has $|\aq| = |\bq|$. It is difficult to
estimate the contribution of $u,d,s$ quarks to $B_D$ perturbatively. However,
one can argue that it is suppressed by the ratio of their current to
constituent (kinematical) masses, and can thus be neglected; exactly the same
argument applies for the contribution of the light quarks to $T_q$,
eq.(\ref{e2}). On the other hand, $I_2$ diverges logarithmically as $m_Q
\rightarrow 0$; since chirality breaking in $B_S$ is due to the LSP mass, not
the quark mass, and the sum of couplings $\asq + \bsq$ remains finite for $m_Q
= 0$, this infrared divergence also survives in eq.(\ref{e5b}). Because the
divergence is only logarithmic, one might be able to estimate the contribution
of the light quarks to $B_S$ by using constituent quark masses in the integral
$I_2$, eq.(\ref{e8b}). However, a consistent treatment of the contribution of
(almost) on--shell light quarks should also include 1--loop contributions to
the LSP--quark scattering amplitude, which is expected to lead to new
spin--independent interactions. In this Letter we focus on LSP--gluon
scattering; we therefore only include the contributions from $c,b,t$ quarks and
squarks in our numerical estimates.

Before presenting our numerical results, we briefly discuss the second and
third term in the effective Lagrangian (\ref{e4}). The coefficient functions
are given by: \ben \label{e10} \beq
B_{1D} &= \gsq \cdot \frac {1}{3} \sum_{\tilde{Q}} \left( \asq - \bsq \right)
m_Q \int_0^1 dx \frac {x^2 (1-x)^2}{D^3}; \label{e10a} \\
B_{1S} &= \gsq \cdot \frac {1}{3} \mc \sum_{\tilde{Q}} \left( \asq + \bsq
\right) \int_0^1 dx \frac {x^3 (1-x)^2}{D^3}; \label{e10b} \\
B_{2S} &= \gsq \cdot \frac {1}{12} \sum_{\tilde{Q}} \left( \asq + \bsq \right)
\int_0^1 dx \frac {x (1-x) (2-x)}{D^2}. \label{e10c} \eeq \een
Unfortunately we do not know how to estimate the hadronic matrix element
$\langle N |$ $F^{\mu a}_{\rho} F^{\rho}_{\nu a}$ $|N \rangle$;\footnote{In the
non--relativistic limit relevant for us, we only need it for $\mu = \nu = 0$;
the contributions with nonzero $\mu$ or $\nu$ are suppressed by the LSP
velocity. $F^{0 a}_{\rho} F^{\rho}_{0 a}$ is the negative of the square of
the chromo--electric field, while $F^{\mu \nu}_a F^a_{\mu \nu}$ is twice the
difference between the squares of the chromo--magnetic and chromo--electric
fields. It seems reasonable to assume that the two matrix elements are of
similar size, but we do not attempt to speculate on their exact ratio.} we do
therefore not give explicit expressions for the integrals appearing in
eqs.(\ref{e10}). Notice, however, that $B_{1D}$ and $B_{1S}$ are suppressed by
a factor $\left( \mc / \msQ \right)^2$ relative to the other contributions;
$B_{1D}$ vanishes for $m_Q \rightarrow 0$, while $B_{1S}$ remains finite in
this limit. On the other hand, $B_{2S}$ is of the same order in $\mc / \msQ$
as $B_S$, and also diverges logarithmically for $m_Q \rightarrow 0$.

In order to demonstrate the numerical importance of the new contributions whose
hadronic matrix element can be estimated reliably, we show in figs.\,2 and 3
the
absolute value of the coefficients $T_q, \ T_{\tilde{q}}, \ B_S$ and $B_D$ of
the effective LSP--gluon interaction Lagrangian, for some typical values of
SUSY parameters. we consider a general model with explicitly, but softly broken
global supersymmetry and minimal particle content, assuming the usual
unification condition \cite{1} between electroweak gaugino masses. We also
assume a common contribution \msqsq\ to all diagonal elements of squark mass
matrices, with \msq\ = 200 GeV, and a common trilinear interaction parameter $A
= \msq$ for all squarks, see eqs.(\ref{em1}). We chose the top quark mass $m_t
= 120$ GeV, the $SU(2)$ gaugino mass $M_2 = 200$ GeV, the pseudoscalar Higgs
mass $m_P = 500$ GeV, and the supersymmetric Higgs(ino) mass parameter $\mu =
-500$ GeV. The last relevant free parameter of the model is the ratio of vevs
\tanb, which we vary along the $x$ axis. The masses of the neutral scalar Higgs
bosons, which enter $T_q$ and $T_{\tilde{q}}$, have been computed including
radiative corrections \cite{17} from the top and bottom sectors, using the
effective potential formalism \cite{18}. For the given choice of parameters,
the LEP Higgs bound is violated for $\tanb < 2$, while for $\tanb > 21$ the
lighter sbottom eigenstate becomes lighter than the lightest neutralino,
leading to a charged LSP.

We see that in this example the contribution of the squark triangle and
bubble diagrams of fig.\,1b is always small. In contrast,
the box and triangle diagrams of fig.\,1c are of the same order of or even
larger than the one from the quark triangle diagrams of fig.\,1a; the exception
is the region of small \tanb, where the $H_2^0$ exchange contributions are
enhanced by the small Higgs mass. For larger values of \tanb, the lighter
scalar Higgs boson does not only get a larger mass, its coupling to the LSP
also decreases rapidly. On the other hand, the coupling of the heavier
scalar Higgs to $b-$quarks is enhanced \cite{14} for large \tanb; for $\tanb
> 8, \ T_q$ is therefore actually dominated by $H_1^0$
exchange in the given example. Both $B_D$ and $B_S$ increase with increasing
\tanb, since \msb\ becomes smaller; the slope is larger for $B_D$, since
$B_S$ is dominated by charm (s)quark contributions, due to the
logarithmic enhancement factor and the larger charge of the charm. (For the
given example, the LSP is mostly a bino.) Notice that the sign of $B_D$ is
opposite that of $B_S$ and $T_q$ here, leading to a vanishing total
contribution
at $\tanb \simeq 11.5$; for larger values of \tanb, $B_D$ dominates. Recalling
that the scattering rate is proportional to the square of the coefficient
plotted in fig.\,2, we conclude that estimates based on the diagrams of
fig.\,1a alone are off by a factor between 0 and 150!

So far we have considered a model with explicitly broken global SUSY. Such a
model can emerge as the low--energy limit of supergravity (SUGRA) models, where
local supersymmetry is broken spontaneously at energies around the Planck
scale; however, since the ``hidden sector'' responsible for SUSY breaking
interacts with the usual gauge and matter superfields only via gravitational
interactions, the size of the SUSY breaking sparticle masses is much smaller
than the amount of SUSY breaking in the hidden sector. In minimal SUGRA models
\cite{1,20} one makes the additional assumption that at the superhigh scale
where SUSY breaking occurs, its effect on the gauge and matter superfields can
be described by just 3 parameters: A common gaugino mass $M$, a common scalar
mass $m_0$, and a common trilinear soft breaking parameter $A$. At lower scales
the degeneracy between scalar masses, and between the three gaugino masses, is
broken due to radiative effects; very often these effects suffice to lead to
spontaneous breaking of the electroweak gauge symmetry at the weak scale. In
such models with radiative gauge symmetry breaking \cite{1,20}, the importance
of the new contributions $T_{\tilde{q}}, \ B_D$ and $B_S$ is usually smaller
than in fig.\,2. This is because in these models one has $\msq \geq 2.5 |M_2|$
due to the contribution of gluino loops to \msq. Moreover, $\mu$ and $m_P$ are
no longer independent parameters, but are determined \cite{15} by the
requirements of electroweak symmetry breaking, for given values of $m_t, \ M,
m_0$ and $A$.

In figs.\,3a,b we have fixed $m_t = 170$ GeV, $m_0 = 150$ GeV,
$M_2 = -120$ GeV, and \tanb = 3, and varied $A$ in the allowed region, which
is determined by the requirement $\mst > \mc$. We present our results as a
function of \mst, which (unlike $A$) is directly measureable.\footnote{Note
that the preliminary CDF bound on squark masses \cite{21}, $\msq \geq 150$ GeV,
does not apply \cite{22} for a single squark flavor; for the given choice of
parameters all squarks except $\tilde{t}_1$ have masses around 300 GeV.}
Any value of \mst\ between zero and 300 GeV can be realized by two different
values for $A$, one above and one below $2.3 m_0$. Since these two solutions
correspond to different values of $\mu$, and thus to different squark mixing
angles and (slightly) different composition of the LSP, they are shown
separately; fig.\,3a (3b) is valid for $A \geq \ (\leq) \ 2.3
m_0$.
%\footnote{
Unlike in ref.\cite{15} we have not assumed the off--diagonal
Higgs mass parameter $\mu_3^2$ to be related to $A$; this allows us to keep
both $A$ and \tanb\ as free parameters. $\mu$ and $m_P$ have been computed
using approximate analytical expressions given in refs.\cite{15,16}.
%}

For $A > 2.3 m_0$ and small \mst, the three new contributions $T_{\tilde{q}}, \
B_D$ and $B_S$ all have the opposite sign as $T_q$, reducing the total
LSP--gluon scattering amplitude by as much as a factor of 2, corresponding to a
reduction of the scattering rate by a factor of 4. However, for $\mst > 100$
GeV or $A < 2.3 m_0$ the new contributions are essentially irrelevant. Their
absolute sizes are suppressed if all squarks are heavy. In addition, there are
cancellations between loops involving different squark eigenstates, as well as
between the three new contributions; e.g., for $A < 0$ and small \mst, there is
an almost perfect cancellation between $B_D$ on the one hand and $B_S$ and
$T_{\tilde{q}}$ on the other. Based on the results of fig.\,3, as well as
further scans of parameter space, we conclude that within the restrictive class
of minimal SUGRA models, the new contributions to the LSP--gluon scattering
amplitude can only be important if \mst\ is not too far above its lower bound;
and due to ``accidental'' cancellations, the sum of the new contributions can
be small even if the stop is light.

Nevertheless, a reduction of the counting rate by a factor of 4 is certainly
not negligible. Moreover, while the minimal SUGRA scenario is attractive due
to its simplicity and predictive power, it need not be realized in Nature;
both the particle physics (naturalness and hierarchy problem) and
cosmological (dark matter problem) motivation for supersymmetry remain
valid in more general models. We have seen above that in such models the
squared LSP--gluon amplitude can differ by a factor between zero and
more than 100 from the conventional estimate based on the diagrams of
fig.\,1a alone.

Unfortunately we were not able to give quantitative estimates of the
additional terms in the effective Lagrangian whose hadronic matrix
elements are not related to the nucleon mass, nor for the contribution of
light quarks and their superpartners. Furthermore, the
LSP--quark scattering amplitude will also have to be computed at
one--loop level before the full spin--independent contribution to the
LSP--nucleon cross section can be calculated reliably. We do therefore not
present any results for cross sections or counting rates in this Letter.
A quantitative understanding of LSP--nucleon interactions is not only
necessary to assess upcoming new direct DM search experiments \cite{23}; it
is also needed for the interpretation of published limits \cite{24} on the flux
of upward going muons in nucleon decay detectors in terms of LSP capture and
annihilation in the Earth or Sun. Our results provide a first step in that
direction; while more work needs to be done, it is already clear that
existing predictions \cite{9,10} for spin--independent LSP--nucleus scattering
will in general have to be modified.

\vspace*{1cm}
{\bf Acknowledgements}
We thank A. Hackfleisch Djouadi and F. Borzumati for useful hints. M.M.N.
greatly appriciates a fellowship provided by the Nishina foundation. This work
is supported in part by the U.S. Department of Energy under contract
No.DE-AC02-76ER00881.

\newpage
\section*{Figure Captions}

\renewcommand{\labelenumi}{Fig. \arabic{enumi}}
\begin{enumerate}
\item  %Fig.1
Feynman diagrams that contribute to LSP -- gluon interactions. Diagrams with
crossed gluon or LSP lines have to be added.

\vspace*{5mm}

\item    %Fig. 2
Results for the absolute value of the coefficients $T_q, \ T_{\tilde{q}}, \
B_D$ and $B_S$ of the effective Lagrangian of eqs.(\ref{e1}) and (\ref{e4}),
in units of ${\rm GeV}^{-3}$. Here we consider a model with explicitly broken
global Supersymmetry.

\vspace*{5mm}

\item    % Fig. 3
Same as fig. 2, but for a minimal Supergravity model with radiative gauge
symmetry breaking. The results in a) and b) correspond to different ranges of
values of the $A$--parameter, and thus also different values of $\mu$ and
$m_P$,
as discussed in the text. Notice that $M_2$ is the SU(2) gaugino mass at the
weak scale, while $m_0$ and $A$ are GUT scale parameters.

\end{enumerate}

\end{document}